%% file: main.tex
\documentclass[sigconf]{acmart}

\setcopyright{acmcopyright}
\copyrightyear{2023}
\acmYear{2023}
\acmDOI{XXXXXXX.XXXXXXX}

\acmConference[EuroMPI 2023]{EuroMPI 2023: The 30th European MPI Users' Group Meeting}{September 11--13,
  2023}{Bristol, UK}

\AtBeginDocument{%
  \providecommand\BibTeX{{%
    \normalfont B\kern-0.5em{\scshape i\kern-0.25em b}\kern-0.8em\TeX}}}

\acmPrice{15.00}
\acmISBN{978-1-4503-XXXX-X/18/06}

\usepackage{algorithmic}
\usepackage[algo2e,ruled,vlined]{algorithm2e}
\SetCommentSty{scriptsize}
\SetKwComment{tcc}{\{}{\}}
\SetKwIF{If}{ElseIf}{Else}{if}{}{else if}{else}{endif}
\SetKwFor{While}{while}{}{endw}
\SetKwFor{ForEach}{for each}{}{endfch}

\SetKwRepeat{Do}{do}{while}

\usepackage{comment}
\usepackage{todonotes}
\usepackage[normalem]{ulem}
\begin{document}

\title{Collective-Optimized FFTs }


\author{Evelyn Namugwanya}
\affiliation{%
  \institution{
  Dept.\@ of Computer Science \\
   Univ.\@ of Tennessee at Chattanooga} 
  \streetaddress{701 E.\@ ML King Jr.\@ Blvd.}
  \city{Chattanooga}
  \state{TN}
  \postcode{37403}
  \country{USA}
}
\email{evelyn-namugwanya@utc.edu}

\author{Amanda Bienz}
\affiliation{%
  \institution{
  Dept.\@ of Computer Science \\
   Univ.\@ of New Mexico} 
  \city{Albuquerque}
  \state{NM}
  \postcode{37403}
  \country{USA}
}
\email{bienz@unm.edu}

\author{Derek Schafer}
\orcid{0000-0001-8438-5144}
\affiliation{%
  \institution{
  Dept.\@ of Computer Science \\
   Univ.\@ of New Mexico} 
  \city{Albuquerque}
  \state{NM}
  \postcode{37403}
  \country{USA}
}
\email{dschafer1@unm.edu}

\author{Anthony Skjellum}
\affiliation{%
  \institution{
Dept.\@ of Computer Science \\
   Univ.\@ of Tennessee at Chattanooga} 
  \streetaddress{701 E.\@ ML King Jr.\@ Blvd.}
  \city{Chattanooga}
  \state{TN}
  \postcode{37403}
  \country{USA}
}
\email{tony-skjellum@utc.edu}

\renewcommand{\shortauthors}{Namugwanya, et al.}

\begin{abstract}


This research proposes a new methodology for optimizing MPI collective communication, specifically  the \texttt{MPI\_Alltoallv} in HPC applications like HeFFTe, a scalable parallel solver for Fast Fourier Transforms (FFTs). Standard implementations of  alltoallv consist  either of sending to a single process and receiving from a single process at each step, bottlenecked by synchronization costs, or initializing all communication at one time, incurring large costs associated with network contention and queue search costs.  The authors present novel methods that eliminate synchronization costs without communicating a large number of messages at once.  

This paper measures the impact of the various alltoallv methods within HeFFTe. Results are analyzed within Beatnik, a Z-model solver that is bottlenecked by HeFFTe and representative of applications that rely on FFTs.  We  evaluate our methodology on  UTC's Epyc cluster. This cluster consists of 16 compute nodes based on the AMD EPYC 7662 128-core processor. 



We made a significant discovery regarding the optimization of OpenMPI \texttt{MPI\_Alltoallv} by utilizing MPI Advance's algorithms. We observed notable reduction in the minimum, maximum, average time and improvements in the scalability.



\end{abstract}


\keywords{Message Passing Interface, collective communication, communication, computation overlap, FFT, \texttt{MPI\_Alltoallv}, Weak scaling.  
}


\maketitle

\newcommand{\PQ}{{$P \times Q$~}}
\newcommand{\QP}{{$Q \times P$~}}

\newcommand{\PQS}{{$P \times Q$}}
\newcommand{\QPS}{{$Q \times P$}}

\input{Sections/Section1}

\input{Sections/Section2}

\input{Sections/Section3}

\input{Sections/Section4}
\input{Sections/Section5}
\input{Sections/Section6}


\begin{acks}
This work was performed with partial support from the National Science
Foundation under Grants Nos.~ 
CCF-1562306, 
CCF-1822191, CCF-1821431, OAC-1923980, OAC-1549812, OAC-1925603,  OAC-2201497, CCF-1918987,  CCF-2151020, and CCF-2151022,  and the U.S. Department of Energy's National Nuclear Security Administration (NNSA) under the Predictive Science Academic Alliance Program (PSAAP-III), Award DE-NA0003966.

Any opinions, findings, and conclusions or recommendations expressed in this material are those of the authors and do not necessarily reflect the views of the 
National Science Foundation, the U.S. Department of Energy's National Nuclear Security Administration.
\end{acks}

\bibliographystyle{ACM-Reference-Format}
\bibliography{references}

\appendix

\input{Sections/Appendix}

\end{document}

%% file: Sections/Section1.tex
\section{Introduction}
\begin{figure}[h]
    \centering
    \includegraphics[width = .4\textwidth]{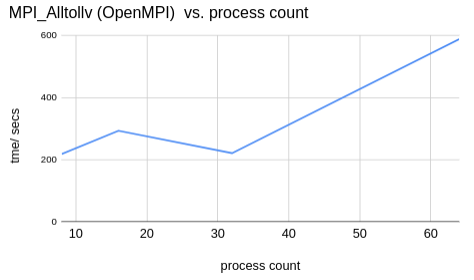}
    \caption{OpenMP's MPI\_Alltoallv's performance vs increase process count}
    \label{fg:op}
\end{figure}

Fast Fourier Transforms (FFTs) are widely used to solve differential equations, filter and compress images, estimate power spectra, and process signals~\cite{brigham1988fast}.  HeFFTe~\cite{HeFFTe}, a state-of-the-art parallel FFT solver, efficiently computes discrete Fourier transforms (via FFT algorithms) on emerging scalable architectures.  As process counts increase, HeFFTe is bottlenecked by a data transpose in which all processes exchange data of varying sizes with all other processes.  Typically, this exchange is implemented with an \texttt{MPI\_Alltoallv}, the cost of which increases with process count.  Each emerging parallel computer brings more compute power through increased core counts.  Therefore, as FFT solvers such as HeFFTe are scaled across increasingly large parallel systems, the alltoallv bottleneck increases in dominance.  The authors present novel implementations of the \texttt{MPI\_Alltoallv}, improving the performance and scalability over existing algorithms.
The \texttt{MPI\_Alltoallv} consists of each process exchanging data with each other process.  As with all collective operations, the cost of the method will increase with process count.  However, in an optimal implementation, this increase should be linear.  With current MPI implementations, the scalability of the \texttt{MPI\_Alltoallv} is far from optimal, as shown in Figure~\ref{fg:op}.

Each parallel computer provides one or a small number of system MPI installations, which are optimized for the given architecture.  While local installations of alternative versions of MPI are possible, the lack of tuning yields suboptimal performance.  Therefore, the authors have published all \texttt{MPI\_Alltoallv} optimizations within MPI Advance, a lightweight library that sits on top of MPI.  These optimizations are available in coordination with most any version of MPI, allowing solvers such as HeFFTe to take advantage algorithmic optimizations while also obtaining the tuned underlying performance of the system MPI installation.  

The contribution of this paper are therefore two-fold.  This paper presents multiple novel implementations of the \texttt{MPI\_Alltoallv} and shows significant improvement to both the performance and scalability of the FFT solver HeFFTe.  Further, the authors present portable communication optimizations that optimize which\-ever version of MPI is tuned for a given system.

As mentioned above, HeFFTe \cite{HeFFTe} is a software library that performs Fast Fourier Transform (FFT) computation on heterogeneous computing systems. The library is designed to target exascale systems, which are computing systems that can perform at least one exaFLOP (floating-point operations per second) of calculations.

Beatnik\footnote{https://github.com/CUP-ECS/beatnik/} is a benchmark for global communication based on Pandya and Shkoller's 3D fluid interface ``Z-Model'' \cite{pandya20223d} in the Cabana/Cajita mesh framework \cite{Cabana2022}. The Beatnik library is bottlenecked by the FFT solves within HeFFTe.
We are optimizing HeFFTe's collective communication, specifically \texttt{MPI\_Alltoallv}. Our goal is to enhance the performance of applications such as Beatnik that rely on FFT efficiency by incorporating improvements to HeFFTe within the Beatnik framework.
In other words, we are optimizing the Beatnik library as representative of applications that rely on FFT performance.

Since the alltoallv dominates the cost of FFT solvers and limits scalability, we developed novel alltoallv implementations that achieve improved performance/scalability over state-of-the-art algorithms.
Our novel alltoallv implementations have demonstrated significant performance improvement in minimum, maximum, average and scalability as compared to traditional approaches.
The novel alltoallv algorithms are derived from  MPI Advance\footnote{https://github.com/mpi-advance}, an MPI extension library authored by some of us and others.
MPI Advance is a collection of MPI extension libraries showcasing new APIs or optimizations of current MPI APIs. 

It is worth noting that our optimization algorithms can be further extended to other all-to-all communication operations, which we leave for future work.

The remainder of the paper is organized as follows: In Section ~\ref{sec:background}, we provide  background, which offers  an overview of some existing knowledge about optimizing MPI collective communication algorithms, along with previous research relevant and related work, to establish the context and setting the foundation for the study. Section~\ref{methodology} details the methodology used in our study. Our results and analysis are presented in Section ~\ref{results}, while Section ~\ref{conclusion} contains the conclusion of our study. Finally, in Section \ref{future work}, we outline potential future work related to optimizing collective communication.

%% file: Sections/Section2.tex
\section{BACKGROUND}
\label{sec:background}
FFT solvers are widely used
to solve mathematical problems related to signal processing, image analysis, data compression, and other fields. The FFT \cite{historyofFFTs} 
solves these systems by efficiently computing the discrete Fourier transform (DFT) of a sequence of complex numbers, reducing the computational complexity from $O(n^2)$ to $O(n log n)$ for a sequence of length n. This makes FFT solvers an essential tool for various applications where DFT needs to be computed rapidly and accurately.  
In parallel, multi-dimensional FFT solvers are bottlenecked by a global transpose, during which all processes are active in \texttt{MPI\_Alltoallv} communication.

There are two standard implementations of the \texttt{MPI\_Alltoallv}, pairwise exchange and non-blocking communication.
\input{Algorithms/pairwise_exchange}
%
The pairwise exchange algorithm, sends to a single process and receives from one process at each step of the algorithm.  For instance, at step $i$, process $p$ sends to process $p-i$ and receives from process $p+i$.  Assuming all processes are ready to communicate, this algorithm minimizes overheads, such as network contention and queue search costs, as only a single message is communicated from each process at once.  However, parallel applications, such as FFTs, often have load imbalances and unsynchronized processes, with some processes working on the \texttt{MPI\_Alltoallv} while others are still computing a previous step.  For instance, assume process $p$ is stepping through the \texttt{MPI\_Alltoallv} algorithm, while process $p-i$ remains in a previous step of the FFT.  Process $p$ initializes a send to process $p-i$ and waits idly, even though process $p-i-1$ may be ready for communication.  This synchronization overhead greatly reduces the performance of the pairwise exchange alltoallv.
\input{Algorithms/nonblocking}
The non-blocking implementation of the \texttt{MPI\_Alltoallv} 
consists of initializing all sends and receives with non-blocking communication, such as \texttt{MPI\_Isend} and \texttt{MPI\_Irecv}.  Then, each process waits for all communication to complete.  This implementation avoids certain synchronization overheads, since all communication is initiated.  Therefore, if process $p-i$ remains in a previous step of the FFT, process $p$ is able to exchange data with every other process while $p-i$ completes previous computations.  However, the non-blocking algorithm incurs large overheads associated with the vast quantity of communication.  All messages are sent through the network at one time, increasing the likelihood of network contention, in which a packet of data needs to traverse a link that is already in use by a separate packet.  Depending on the routing algorithm, the packet may sit idly until the link is free of contention, adding large delays to message routing.  Further, large message counts incur significant queue search overheads.  Each process posts \texttt{MPI\_Irecv} calls for every message it expects to receive.  Once a message arrives, all posted receives are searched to find a match.  As the number of messages grows, this queue search cost incurs large performance overheads.
We used Tau \cite{TAU} and Caliper \cite{Caliper1} to profile Beatnik, specifically focusing, of course, on \texttt{MPI\_Alltoallv}.
We modified the HeFFTe library and replaced  OpenMPI's alltoallv with MPI Advance’s alltoallv.
We configured Spack \cite{spack} to link to our edited version of HeFFTe, which is called by the Beatnik benchmark.

\section{Other related work}
A substantial body of related work has been done to improve the performance of collective operations, such as the \texttt{MPI\_Alltoallv}.
A large number of optimizations rely on architecture-awareness to minimize communication costs.
For instance, one common theme in collective optimizations is minimizing the message sizes and counts communicated between sets of nodes, because intra-node communication often greatly outperforms inter-node.  Hierarchical collectives achieve this minimum by gathering all data to one or a small number of leader processes per node before performing the collective between only the leader processes.  Finally, all results are broadcast locally within each node~\cite{KaronisHierColl2000, TraffHierAllgather2006, KandallaMultliLeaderHier2009}.  Multi-lane collectives further optimize large collective operations, with each process per node communicating an equal but minimal amount of data~\cite{TraffMultilane2020}.  Finally, locality-aware collectives minimize the number of inter-node messages, with each process per node communicating with a separate subset of nodes~\cite{BienzLocAwareBruck2022, BienzLocAwareAllreduce2019}.

Topology-aware optimizations also rely on architecture-aware\-ness, minimizing communication for a given interconnect.  There are two main approaches for topology-aware collectives: (1) to remap data to processes and (2) create new underlying algorithms; both of these approaches minimize the number of network links traversed during communication~\cite{MirsadeghiTopoAwareRRCollectives2016, MaHierKNEMTopoAware2012, PatarasukTreeAllreduce2007}.

While the Bruck algorithm was created for \texttt{MPI\_Alltoall} algorithms with small data sizes, it has recently been extended for variable data sizes within the \texttt{MPI\_Alltoallv}~\cite{variableBruck}.  However, the authors are unaware of general architecture-agnostic optimizations for large \texttt{MPI\_Alltoallv} algorithms.

 Benson et al\hbox{.} evaluated the performance of \texttt{MPI\_Allgather} in MPICH 1.2.5 on a Linux cluster and found that the implementation of MPICH improved the performance of allgather compared to previous versions by using a recursive doubling algorithm. They also developed a dissemination Allgather based on the dissemination barrier algorithm that takes log2 p stages for any values of p. The authors experimentally evaluated MPICH's Allgather and their implementations on a Linux cluster of dual-processor nodes using both TCP over FastEthernet and GM over Myrinet. They found that variations of the dissemination algorithm perform best for both large and small messages on Myrinet, but that the dissemination allgather algorithm performs poorly when using TCP because data is not exchanged in a pair-wise fashion. Therefore, they recommended the dissemination allgather for low-latency switched networks \cite{AComparisonofMPICHAllgatherAlgorithmsonSwitchedNetworks}.

Netterville et al\hbox{.} designed a parameterized visualization system to illustrate the technical details of the spread-out algorithm and the Bruck algorithm, as well as the selection decision tree between them for \texttt{MPI\_Alltoall}, which is one of the most widely used collective routines. The system is intended for HPC users who would like to learn or teach about the detailed implementation of \texttt{MPI\_Alltoall}, which can greatly cut the learning curve.
The authors also plan to extend their visualization system to incorporate modified versions of the Bruck algorithm, which take advantage of the multi-node-multi-core architecture. Additionally, they intend to enable their system to visualize other collective routines in the future, such as \texttt{MPI\_Allgather}, \texttt{MPI\_Alltoallv}, and \texttt{MPI\_Allreduce}, which are commonly used in HPC applications and use a combination of several algorithms and a decision tree to select between them 
\cite{visualguide}.

%% file: Algorithms/pairwise_exchange.tex
\begin{algorithm2e}[ht!]
  \DontPrintSemicolon%
  \KwIn{$p$\tcc*{process id}
        $n$\tcc*{number of processes}
        $\texttt{args}$\tcc*{arguments passed to MPI\_Alltoallv}
        }

  \BlankLine%
  
  \For{$i\gets0$ \KwTo $n$}{
      $p_{\texttt{send}} = p + i \mod n$\;
      $p_{\texttt{recv}} = p + n - i \mod n$\;
      Send message to $p_{\texttt{send}}$ and receive message from $p_{\texttt{recv}}$\;
  }
  
    \caption{Pairwise Exchange}\label{alg:pairwise_exchange}
\end{algorithm2e}

%% file: Algorithms/nonblocking.tex
\begin{algorithm2e}[ht!]
  \DontPrintSemicolon%
  \KwIn{$p$\tcc*{process id}
        $n$\tcc*{number of processes}
        $\texttt{args}$\tcc*{arguments passed to MPI\_Alltoallv}
        }

  \BlankLine%
  
  \For{$i\gets0$ \KwTo $n$}{
      $p_{\texttt{send}} = p + i \mod n$\;
      $p_{\texttt{recv}} = p + n - i \mod n$\;
      Initialize non-blocking send to $p_{\texttt{send}}$\;
      Initialize non-blocking receive from $p_{\texttt{recv}}$\;
  }

  Wait for all sends and receives to complete\;
  
    \caption{Non-Blocking}\label{alg:nonblocking}
\end{algorithm2e}

%% file: Sections/Section3.tex
\section{METHODOLOGY}
\label{methodology}
This section provides an in-depth explanation of our methodology, thereby offering  insights into our approach. 
In order to validate our hypothesis---that we can improve \texttt{MPI\_Alltoallv} performance and thereby improve HeFFTe and Beatnik---we have developed three novel extensions to alltoallv:  multi-pair exchange, multi-pair nonblocking exchange, and multi-pair test exchange, which are 
described and analyzed throughout this section.

\label{Implementaiton}
 We describe functionality of all versions of MPI Advance`s alltoallv algorithms used for profiling in this paper. Our experiments, results, and conclusions are based on these algorithms.

As noted above, and based on a formal specification, During an alltoallv pairwise communication, each process normally sends a message to every other process in the communicator. The amount of data sent from each process to each other process can be different, which is explains the ``v" in ``alltoallv." This is a blocking operation.  The inherent blocking nature of the operation necessitates processes to remain idle until all message exchanges are completed, which can result in decreased efficiency and wasted time. To overcome these challenges, one can explore alternative methods like non-blocking or optimized algorithms to enhance the performance of alltoallv pairwise communication.
Since the non-blocking algorithm allows processes to initiate communication and proceed without waiting for completion, it allows processes to continue searching the queue while waiting for messages from other processes which results into a long queue search thus increased complexity.

We address this challenge with a multi-pair exchange, which sends to \texttt{stride} different processes at once.  While this allows for more flexibility than the standard pairwise exchange, it still results in synchronization costs.  If, one of the many \texttt{stride} processes is not ready, the communicating waits idly.  The multi-pair waitany exchange further improves upon this by waiting for any one of the \texttt{stride} processes to finish and then communicating the next, allowing for the algorithm to continue even if some processes have not yet started their portion of the \texttt{MPI\_Alltoallv}.

Further, we considered the implementation of multi-pair test exchange; When utilizing multi-pair nonblocking exchange, the program execution pauses until any of the communication operations (exchanges) is successfully executed. After the completion of any exchange, the program can then proceed to the subsequent step or operation. This indicates that the program remains inactive until at least one exchange is finished, allowing for further progression.  However, the multi-pair test exchange allows uninterrupted program execution by monitoring ongoing exchanges and promptly returning the status of the first completed exchange, if any, without blocking or halting.



\subsubsection{Multi-pair blocking exchange}
\label{sec:opt_nonblocking_pairwise_alg}
\input{Algorithms/pairwise_nonblocking}
 Just like the name suggests, multi-pair blocking exchange algorithm uses uses waitall to wait on multiple pairs of processes that are exchanged with at each step. Multi-pair blocking exchange improves performance over pairwise exchange because it communicates with a number of processes at once, so if one process is not ready, it is still able to communicate with all others in a group. Multi-pair blocking exchange also improves performance over non-blocking pairwise exchange because only `stride' messages are communicated , there by reducing network contention and queue search costs over standard non-blocking. 
 Multi-pair blocking exchange operates similarly to a non-blocking version of the alltoallv pairwise communication operation. 
 It uses the pairwise pattern to communicate between processes. The algorithm also uses non-blocking calls to overlap computation with communication. 
 
 Similar to alltoallv pairwise algorithm, this algorithm has a for loop that iterates over all processes in the communicator (with rank values ranging from 0 to num\_procs-1), and for each process, it calculates the ranks of the processes to which it will send and receive data. The send and receive positions in the sendbuf and recvbuf arrays are also calculated based on the send and receive displacements and sizes (stored in sdispls, rdispls, sendcounts, and recvcounts arrays).
The \texttt{MPI\_Isend} and \texttt{MPI\_Irecv} functions are used to initiate non-blocking sends and receives of data between the two processes. The requests array stores the requests for these communications.
Unlike alltoallv pairwise and alltoallv non-blocking pairwise algorithms, this algorithm has a tuning parameter, stride. We could regulate this parameter to 5, 10, or 15 depending on the number of processes we are interested in waiting on at a time and this can yield a speedup.

\subsubsection{Multi-pair nonblocking exchange}
\input{Algorithms/pairwise_nonblocking_waitany}

The multi-pair nonblocking exchange algorithm operates in a similar manner to the multi-pair blocking exchange. However, it differs in that it does not wait for all `stride' processes to finish before proceeding. Instead, it waits for one exchange to be completed. By doing so, the multi-pair non-blocking exchange algorithm enhances performance compared to the multi-pair blocking exchange. This improvement is achieved by allowing progress to be made in initializing additional communication, even if some of the stride processes involved in the exchange are not yet ready. As soon as one of the 'stride' messages is completed, this approach can continue its execution, thereby optimizing overall efficiency.

The algorithm declares all necessary variables. 
The algorithm enters a for-loop and iterates over all stride-based elements (a number of elements in the tuning parameter).
The algorithm determines the source, send\_proc (rank - I), and destination, send\_proc process (rank + i).
Next, the algorithm initializes \texttt{MPI\_Isend} and posts a matching Irecv.
The algorithm uses an if condition to wait for all processes if the number of `stride' processes exceeds the total number of processes.

If the number of `stride' processes is less than the total number of processes, which applies to most cases, then the algorithm 
enters a while loop, loops continuously, and waits on any processes to complete. The algorithm checks for any valid processes and uses the MPI-index (idx) to determine if the process is a source process or a destination process. 

If there are send processes or source processes, the process index is even, then that's a sending process, or it's a receiving process.
The algorithm calculates the sending and receiving positions, displacements accordingly and adjusts the send and receive buffers.
The algorithm calls non-blocking Send on a send process and increments the send process index to go to the next send process.
The Algorithm calls non-blocking Recv on a receive process
and increments the receive process index to go to the next receive process.
Finally the algorithm frees all memory requests and returns 0.

\subsubsection{Multi-pair test exchange}
\input{Algorithms/pairwise_nonblocking_testany}
This algorithm works similarly to the previous algorithm.  
However, in the case of multi-pair nonblocking exchange, the program execution waits for any one of the communication operations (exchanges) to complete. Once any of the exchanges is finished, the program proceeds to the next step or operation. This implies that the program halts until at least one exchange is completed before it can continue.
On the other hand, multi-pair test exchange does not block or halt the program's execution while waiting for an exchange to complete. Instead, it checks the status of all the ongoing exchanges and immediately returns the index of the first completed exchange, if any. If no exchange has completed yet, it returns a special value indicating that no communication operation has indeed finished. This allows the program to continue executing other tasks or operations without waiting for any exchange to complete.

The algorithm enters a for-loop and iterates over all stride-based elements, number of times,
which should be less than the number of processes.
Within the loop, the code calculates the ranks of the processes to which data is to be sent and received. It then calculates the positions of the send and receive buffers and sends asynchronous non-blocking send and receives requests using the \texttt{MPI\_Isend} and \texttt{MPI\_Irecv} functions, respectively. Each request is stored in an array of requests, and the ctr variable is incremented by 2 with each iteration of the loop.
After the loop, the code checks if the value of nb\_stride is greater than or equal to num\_procs. If it is, it means that all the data can be sent and received in a single iteration of the loop. In this case, the code waits for all the requests to complete using the \texttt{MPI\_Waitall} function and then frees the memory allocated for the requests array before returning 0.
If the value of nb\_stride is less than num\_procs, which is the case most of the time, the code enters a while loop that continuously tests for the completion of any of the send or receives requests using the \texttt{MPI\_Testany} routine. If a request has been completed, the code checks if it was a send or receive request and sends another request of the same type to the next process if there are any remaining processes from which to send or receive data. The loop continues until all the data has been sent and received, at which point the memory allocated for the requests array is freed before the code returns 0.

%% file: Algorithms/pairwise_nonblocking.tex
\begin{algorithm2e}[ht!]
  \DontPrintSemicolon%
  \KwIn{$p$\tcc*{process id}
        $n$\tcc*{number of processes}
        $\texttt{stride}$\tcc*{Number of non-blocking messages at a time}
        $\texttt{args}$\tcc*{arguments passed to MPI\_Alltoallv}
        }

  \BlankLine%
  
  \For{$i\gets0$ \KwTo $n$}{
      $p_{\texttt{send}} = p + i \mod n$\;
      $p_{\texttt{recv}} = p + n - i \mod n$\;
      Initialize non-blocking send to $p_{\texttt{send}}$\;
      Initialize non-blocking receive from $p_{\texttt{recv}}$\;

    \uIf {$i \mod \texttt{stride} == 0$}
    {
      Wait for $\texttt{stride}$ sends and receives to complete\;
    }
  }
  
    \caption{multi-pair blocking exchange}\label{alg:pairwise_nonblocking}
\end{algorithm2e}

%% file: Algorithms/pairwise_nonblocking_waitany.tex
\begin{algorithm2e}[ht!]
  \DontPrintSemicolon%
  \KwIn{$p$\tcc*{process id}
        $n$\tcc*{number of processes}
        $\texttt{stride}$\tcc*{Number of non-blocking messages at a time}
        $\texttt{args}$\tcc*{arguments passed to MPI\_Alltoallv}
        }

  \BlankLine%
  
  \For{$i\gets0$ \KwTo $\texttt{stride}$}{
      $p_{\texttt{send}} = p + i \mod n$\;
      $p_{\texttt{recv}} = p + n - i \mod n$\;
      Initialize non-blocking send to $p_{\texttt{send}}$\;
      Initialize non-blocking receive from $p_{\texttt{recv}}$\;
   }

   \While{idx != MPI\_UNDEFINED}{
       $\texttt{idx} \leftarrow$ MPI\_Waitany of $stride$ messages\;
        \uIf {$\texttt{idx}$ was a send and $p_{\texttt{send}} \neq p$}
        {
            Initialize non-blocking send to $p_{\texttt{send}}$\;
            $p_{\texttt{send}}$ = $p_{\texttt{send}} + 1 \mod n$\;
        }
        \uElseIf {$\texttt{idx}$ was a receive and $p_{\texttt{recv}} \neq p$}
        {
            Initialize non-blocking receive from $p_{\texttt{recv}}$\;
            $p_{\texttt{recv}}$ = $p_{\texttt{recv}} - 1 \mod n$\;
        }
    }

    \caption{multi-pair nonblocking exchange}\label{alg:pairwise_nonblocking_waitany}
\end{algorithm2e}

%% file: Algorithms/pairwise_nonblocking_testany.tex
\begin{algorithm2e}[ht!]
  \DontPrintSemicolon%
  \KwIn{$p$\tcc*{process id}
        $n$\tcc*{number of processes}
        $\texttt{stride}$\tcc*{Number of non-blocking messages at a time}
        $\texttt{args}$\tcc*{arguments passed to MPI\_Alltoallv}
        }

  \BlankLine%
  
  \For{$i\gets0$ \KwTo $\texttt{stride}$}{
      $p_{\texttt{send}} = p + i \mod n$\;
      $p_{\texttt{recv}} = p - i \mod n$\;
      Initialize non-blocking send to $p_{\texttt{send}}$\;
      Initialize non-blocking receive from $p_{\texttt{recv}}$\;
   }

   \While{idx != MPI\_UNDEFINED}{
       \While{($\texttt{idx} \leftarrow$ MPI\_Testany $stride$ messages) == \texttt{False}}{}
        
        \uIf {$\texttt{idx}$ was a send and $p_{\texttt{send}} \neq p$}
        {
            Initialize non-blocking send to $p_{\texttt{send}}$\;
            $p_{\texttt{send}}$ = $p_{\texttt{send}} + 1 \mod n$\;
        }
        \uElseIf {$\texttt{idx}$ was a receive and $p_{\texttt{recv}} \neq p$}
        {
             Initialize non-blocking receive from $p_{\texttt{recv}}$\;
             $p_{\texttt{recv}}$ = $p_{\texttt{recv}} - 1 \mod n$\;
        }
    }

    \caption{Pairwise Non-Blocking Testany}\label{alg:pairwise_nonblocking_testany}
\end{algorithm2e}

%% file: Sections/Section5.tex
\section{Results and Analysis}
\label{results}

 We performed experiments on a cluster called Epyc, specifically the UTC  AMD cluster, which has 16 compute nodes powered by the EPYC 7662 128 core processor. The production part of the cluster has a total of 2,048 cores.
 Our goal was to evaluate the performance of all \texttt{MPI\_Alltoallv} algorithms described in Section ~\ref{methodology}. 

We assessed the efficiency of various algorithms by testing seven different setups, measuring the minimum, maximum, and average times aggregated across all processes for different versions of alltoallv collective communication;
\begin{itemize}
  \item \textbf{Alltoallv Pairwise (Algorithm~\ref{alg:pairwise_exchange})}
  \item \textbf{Non-blocking Alltoallv (Algorithm ~\ref{alg:nonblocking})}
  
  \item \textbf{Multi-pair blocking exchange (Algorithm~\ref{alg:pairwise_nonblocking})}
  
  \item \textbf{Multi-pair nonblocking exchange (Algorithm~\ref{alg:pairwise_nonblocking_waitany})}
  
  \item \textbf{Multi-pair Test Exchange (Algorithm~\ref{alg:pairwise_nonblocking_testany})}

 \item \textbf{OpenMPI Alltoallv (default implementation selected by OpenMPI 4.1.0}
\end{itemize}


In our study, we carried out seven tests, keeping the problem size and grid size constant while varying the number of processes and nodes participating in each operation. We conducted multiple tests using different versions of MPI Advance's alltoallv algorithms, and OpenMPI's Alltoallv, considering one to eight nodes and adjusting the number of processes accordingly. To accurately measure the duration of each \texttt{MPI\_Alltoallv} operation in HeFFTe, we utilized profiling tools like Tau and Caliper.
To observe the variations of our measurements, we repeated each experiment five times and we took the best time for each experiment. 
The results of the best experiments for each algorithm out of five times are presented in graphs in this Section. 

Overall, our experiments demonstrate the effectiveness of using MPI Advance's Multi-pair nonblocking exchange for faster MPI collective communication operations in HeFFTe. Almost all MPI Advance's alltoallv algorithm perform better than OpenMPI's Alltoallv. These findings can help developers to further optimize collective communication in FFTs and other applications which use MPI collective communication. These findings can also be used by Open MPI developers to review the choice of the default algorithm for FFTs or a given message size.

The experiments showed that the performance of the collective communication algorithms varies significantly depending on the number of processes, nodes and the grid size which scales up with the number of processes. 
\subsection{Analysis}
In this simulation, we did a performance analysis of the Beatnik benchmark as a representative for FFT applications that call HeFFTe. The simulation used a fixed problem size with a grid size of 4,096, which was scaled up by varying the number of processes from 8 to 64. We increased the grid size linearly with the number of processes, using the square root factor of the number of processes.

For instance, when we used 64 processes, the mesh dimensions were 32,768 by 32,768, which is equivalent to $4,096 times 8$ by $4,096 \times 8$. When we used 32 processes, the square root was about 5.66, resulting in a mesh dimension of 23,170 by 23,170. As the number of processes increased, the problem size increased as well.

\begin{figure}[h]
    \centering
    \includegraphics[width = .4\textwidth]{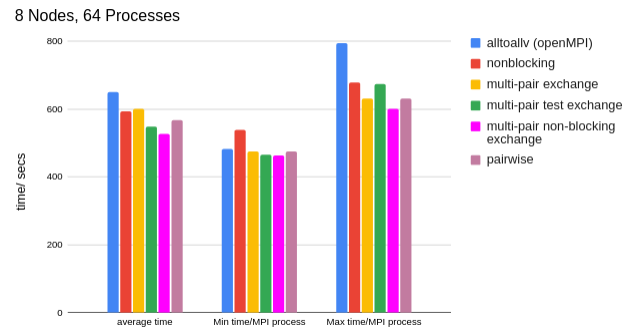}
    \caption{Average, Minimum and Maximum time taken by different versions of OpenMPI alltoallv vs\hbox{.} MPI Advance's alltoallv algorithms}
    \label{fg:amm}
\end{figure}

\begin{figure}[h]
    \centering
    \includegraphics[width = .4\textwidth]{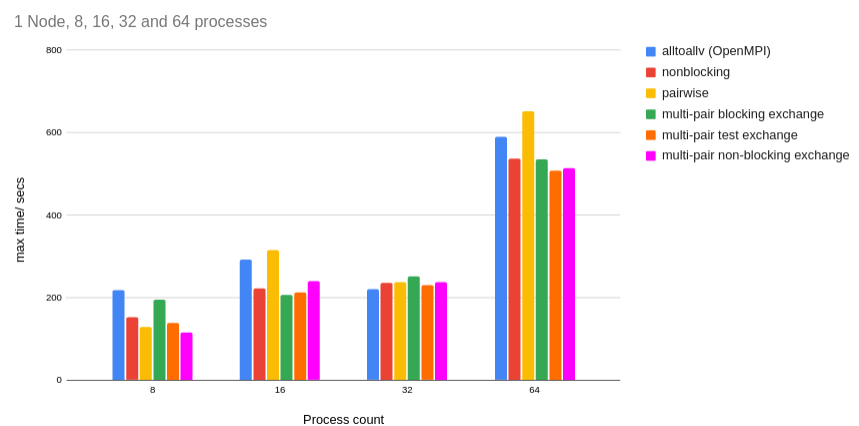}
    \caption{Maximum time taken by different versions of Open MPI's alltoallv vs\hbox{.} MPI Advance's alltoallv algorithms}
    \label{fg:a}
\end{figure}

MPI Advance’s alltoallv pairwise perform better with smaller grid sizes, problem sizes, while MPI Advance’s multi-pair non-blocking exchange, and multi-pair test exchange perform well with large or small grid sizes. This was observed in the results section, (e.g., Figure~\ref{fg:amm}, Figure~\ref{fg:a}) where we compared the performance of different node and process configurations.

In Figure~\ref{fg:amm}, we present data showcasing the average, maximum, and minimum time taken by Open MPI's alltoallv algorithm and the optimized MPI Advance's alltoallv algorithms. Our performance observations were conducted on these algorithms using 8 nodes and 64 processes, with a fixed grid size that scales up according to the number of processes. The colors in the figure represent the different algorithms, and the x-axis represents the average, minimum, and maximum time counts.
Overall, some of our novel algorithms outperformed the standard \texttt{MPI\_Alltoallv}, significantly reducing the average time taken during communication operations across all processes. Notably, the multi-pair nonblocking exchange algorithm emerged as the top performer in terms of reducing average, minimum, and maximum time, particularly  as we increased the number of nodes.

In Figure~\ref{fg:a}, we  display the maximum time taken by different versions of OpenMPI's alltoallv and MPI Advance's alltoallv algorithms. We conducted performance observations on all algorithms with varying numbers of processes on one node, with a fixed grid size that scales up with the number of processes. The colors in the figure represent different algorithms and we have process count along the x-axis.
Our observations indicate that all algorithms perform better with a smaller number of processes (8, 16, 32) than with a higher number of processes. MPI Advance's algorithms, such as multi-pair nonblocking exchange and multi-pair test exchange, perform well for both small and large numbers of processes. This can be seen for 8, 16, and 64 processes, as compared to Open MPI's Alltoallv and other versions of MPI Advance algorithms.
For 1 node and 32 processes, Open MPI alltoallv outperformed all the algorithms. However, for larger numbers of processes (64 processes), multi-pair test exchange outperforms Open MPI's alltoallv and all other versions of alltoallv in MPI Advance. These results suggest that as we increase the number of processes and scale up the grid size and problem size, all MPI Advance's alltoallv algorithms perform well, except for alltoallv pairwise, which performs best with a smaller number of processes, smaller problem size, smaller grid size. Notably, multi-pair nonblocking exchange and multi-pair test exchange's performance stand out as compared to the standard Open MPI and rest of MPI-Advance's alltoallv algorithms.

\begin{figure}[h]
    \centering
    \includegraphics[width = .4\textwidth]{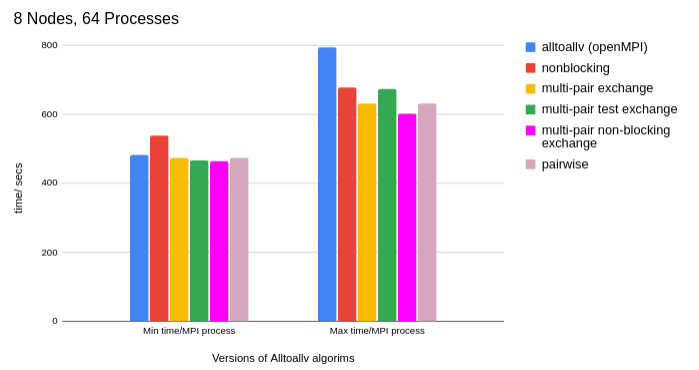}
    \caption{Maximum and Minimum time taken by different versions of  Alltoallv on 8 Nodes, 64 Processes}
    \label{fg:c}
\end{figure}

Figure~\ref{fg:c} shows 
on all algorithms using 64 processes across eight nodes. 
Our analysis confirms that the MPI Advance's multi-pair nonblocking exchange algorithm outperforms OpenMPI's alltoallv and all other versions of alltoallv in MPI Advance. This result highlights the efficacy of the MPI Advance's multi-pair nonblocking exchange algorithm for use in FFTs application like beatnik and HeFFTe.

\begin{figure}[h]
    \centering
    \includegraphics[width = .4\textwidth]{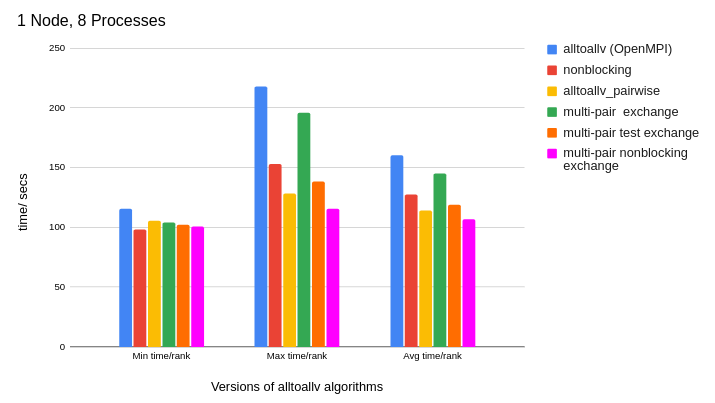}
    \caption{Average, Maximum and Minimum time taken by different versions of OpenMPI alltoallv vs\hbox{.} MPI Advance's alltoallv algorithms}
    \label{fg:ee}
\end{figure}
When examining a smaller scale scenario with 1 node and 8 processes in Figure ~\ref{fg:ee}, the multi-pair non-blocking exchange algorithm achieved a reduction in the average execution time across all ranks by a certain number of seconds. 

Additionally, this algorithm decreases the maximum time and the minimum time, as well as reducing the average time across all processes. 
In terms of reducing the average time taken during communication operations across all processes, our novel algorithms from MPI Advance consistently outperform the standard \texttt{MPI\_Alltoallv}. Notably, the multi-pair nonblocking exchange algorithm shines in its ability to reduce the average, minimum, and maximum time. 
This test provides evidence that the multi-pair non-blocking exchange algorithm outperforms other algorithms across a range of resource variations, from 1 to 8 nodes.

%% file: Sections/Section6.tex
\section{Conclusion}
\label{conclusion}
In conclusion, we achieved our goal of making HeFFTe faster by improving \texttt{MPI\_Alltoallv} communication.
This study proposed a new methodology to optimize MPI collective communication, with HeFFTe used as an important exemplar. 
The method involves profiling the Beatnik library with Tau and Caliper, focusing on \texttt{MPI\_ALLtoallv} to identify performance characteristics. 
The HeFFTe library was modified by replacing the \texttt{MPI\_Alltoallv} routine with optimized algorithms from the MPI Advance library, minimizing the number of messages and data movement between processes, while overlapping communication with computation. This results in a significant speedup in the collective routine.
The observations suggest that all MPI Advance's alltoallv algorithms perform well as we increase the number of processes and scale up the grid size and problem size, except for alltoallv pairwise. Notably, multi-pair nonblocking exchange's performance stands out, outperforming Open MPI's alltoallv  and all other versions of alltoallv in MPI Advance for larger numbers of processes (64 processes) and some small cases.

In six out of seven test scenarios, multi-pair nonblocking exchange performs better than Open MPI's alltoallv. 
These findings provide  effective ways for optimizing MPI collective communication and have the potential to enhance the performance of various parallel computing applications in the field of HPC.

\section{Future Work}
\label{future work}
There are many extensions to this work that have the potential to further improve the performance and scalability of collective operations such as the \texttt{MPI\_Alltoallv}.  As a large majority of supercomputers are heterogeneous and applications are often accelerated on the available GPUs, FFT solvers and the associated alltoallv methods should be optimized for data movement between GPUs.  For instance, HeFFTe developers have noted 2x-3x performance gains \cite{HeFFTe} in moving from CPU to GPU.  GPU-aware \texttt{MPI\_Alltoallv} optimizations will introduce several new aspects of complexity: (1) the ability of the algorithms to move data asynchronously between GPU memories as a function of the network architecture and MPI implementation, (2) the ability to partition certain operations to enhance overlap of communication and computation (e.g., partitioned collective algorithms \cite{holmes2021partitioned}), (3) new bottlenecks associated with the completion of GPU kernels vs.\@ the initiation of transfers, and other possible effects involving greater message rate because of enhanced computing speed involving the GPUs.  

Similar communication optimizations can be explored for other collective operations.  Further, the combination of the methods presented in this paper and locality-aware aggregation or hierarchical methods can be combined to further optimize methods such as the \texttt{MPI\_Alltoallv}.
